%% file: main.tex
\pdfoutput=1
\documentclass[a4paper,11pt]{article}
\usepackage[tableposition=above,font=small,labelfont={small,bf},labelsep=period]{caption}
\usepackage{jcappub_mod}

\usepackage[english]{babel}
\usepackage[T1]{fontenc}
\usepackage{lmodern}
\usepackage{upgreek}
\usepackage{pifont}
\usepackage{fontawesome}
\usepackage{relsize}
\usepackage{microtype}
\usepackage{xspace}
\usepackage{soul}
\usepackage{textcomp}
\usepackage{amsmath}
\usepackage[noabbrev]{cleveref}
\crefname{equation}{eq.}{eqs.}
\Crefname{equation}{Equation}{Equations}
\crefname{enumi}{step}{steps}
\Crefname{enumi}{Step}{Steps}

\usepackage[dvipsnames]{xcolor}
\usepackage{graphicx}
\usepackage{booktabs}
\usepackage{tabularx}
\usepackage{enumitem}
\usepackage{makecell}
\usepackage[export]{adjustbox}
\usepackage{soul}

\usepackage{siunitx}
\DeclareSIUnit\astrounit{au}
\DeclareSIUnit\year{yr}
\usepackage[version=4]{mhchem}

\usepackage{tikz}
\usetikzlibrary{positioning,matrix,calc}

\usepackage[sort&compress,numbers]{natbib}
\input{include/journals}

\definecolor{Blue}{RGB}{0,0,122}
\definecolor{Red}{RGB}{173,42,26}
\hypersetup{colorlinks=true, linktoc=all, linkcolor=Blue, citecolor=Red, urlcolor=Blue}

\input{include/definitions}
\newcommand{\updated}[1]{#1}

\title{\fontsize{23}{\baselineskip}\selectfont Axion Helioscopes as Solar Thermometers}

\author[a,b]{Sebastian Hoof,}
\author[c]{Joerg Jaeckel,}
\author[\,c]{and Lennert J.\ Thormaehlen}

\affiliation[a]{Dipartimento di Fisica e Astronomia ``Galileo Galilei,''
{Universit\`a} degli Studi di Padova,\\ Via F.\ Marzolo 8, 35131 Padova, Italy}
\affiliation[b]{Istituto Nazionale di Fisica Nucleare -- Sezione di Padova,\\
Via F.\ Marzolo 8, 35131 Padova, Italy}
\affiliation[c]{Institut f\"{u}r Theoretische Physik, Universit\"{a}t Heidelberg,\\ Philosophenweg 16, 69120 Heidelberg, Germany}
\emailAdd{hoof@pd.infn.it}
\emailAdd{jjaeckel@thphys.uni-heidelberg.de}
\emailAdd{l.thormaehlen@thphys.uni-heidelberg.de}

\abstract{
Axions, if discovered, could serve as a powerful new messenger for studying astrophysical objects.
In this study we show how the Sun's spatial and spectral ``axion image'' can be inverted to infer the radial dependence of solar properties in a model-independent way.
In particular, the future helioscope IAXO may allow us to accurately reconstruct the Sun's temperature profile $T(r)$ in the region up to about 80\% (40\%) of the solar radius for an axion-photon coupling $g_{a\gamma\gamma}$ of \SI{6e-11}{\GeV^{-1}} (\SI{e-11}{\GeV^{-1}}).
The statistical fluctuations in the photon data lead to a median precision of better than 10\% (16\%) in this region, and the corresponding median accuracy was better than 4\% (7\%).
While our approach can simultaneously infer the radial profile of the Debye scale $\kappa_\text{s}(r)$, its weaker connection to the axion production rate leads to median accuracy and precision of worse than 30\% and 50\%, respectively.
We discuss possible challenges and improvements for realistic setups, as well as extensions to more general axion models.
We also highlight advantages of helioscopes over neutrino detectors.~\href{https://github.com/sebhoof/SolarAxionFlux}{\faGithub}
}

\begin{document}
\maketitle
\flushbottom

\section{Introduction}
Numerous experimental searches~\cite{1801.08127} are currently underway or planned to look for QCD~axions~\cite{Peccei:1977hh,Peccei:1977ur,Wilczek:1977pj,Weinberg:1977ma} and axion-like particles~(ALPs)~\cite{Kim:1986ax,1002.0329}.
Axions are hypothetical particles, which could solve many of the open questions in physics.
These include the Strong~CP problem~\cite{Peccei:1977hh,Peccei:1977ur}, dark matter~\cite{Preskill:1982cy,Abbott:1982af,Dine:1982ah,Turner:1983he,Turner:1985si,1201.5902}, and anomalous observations in a variety of astrophysical environments~\cite[e.g.][]{0807.4246,1201.4711,1512.08108,1708.02111}.

An important class of axion experiments are helioscopes~\cite{Sikivie:1983ip,Sikivie:1985yu,vanBibber:1988ge}, which track the position of the Sun and use strong magnetic fields to convert solar axions into X-ray photons.
Previous helioscope campaigns~\cite{Lazarus:1992ry,hep-ex/9805026,astro-ph/0204388,hep-ex/0411033,hep-ex/0702006,0806.2230,0906.4488,1106.3919,1302.6283,1307.1985,1503.00610,1705.02290} placed limits on ALP-photon interactions for sub-meV axions, with the strongest limit of $\gagg < \SI{0.66e-10}{\GeV^{-1}}$~(at 95\% confidence level) coming from the CAST experiment~\cite{1705.02290}.
The upcoming helioscope IAXO~\cite{1401.3233,1904.09155,2010.12076} is expected to improve on the CAST sensitivity by more than one order of magnitude.

In addition to the potential discovery of axions, IAXO may even determine the axion's mass and couplings~\cite{1811.09278,1811.09290}.
Axions could also be used to study solar properties such as metal abundances~\cite{1908.10878}, macroscopic magnetic fields inside the Sun~\cite{2005.00078,2006.10415}, or to distinguish different QCD axion or solar models~\cite{2101.08789}.

In this study, we continue to explore the capabilities of helioscopes as a tool for studying solar properties.
Specifically, we demonstrate how helioscopes can leverage pixel-based detectors and the excellent angular resolution of X-ray optics to determine the solar temperature and Debye screening scale in different layers of the Sun's interior.

In \cref{sec:methodology} and \cref{app:reconstruction}, we outline the methodology for inferring solar properties as a function of distance from the solar core using the axion's energy-averaged interaction rates.
In \cref{sec:case_study}, we present a case study that demonstrates the feasibility of our approach using simulated IAXO data.
\Cref{sec:comments} provides a comprehensive discussion of our assumptions, potential practical challenges, and draws connections to neutrino experiments and inversion methods previously employed in solar physics.
Finally, in \cref{sec:conclusions}, we summarise our key findings and close with a few concluding remarks.

\section{Reconstructing solar properties from helioscope data}\label{sec:methodology}

Depending on their interactions, QCD axions and ALPs can be created inside the Sun through various processes, which we extensively reviewed in ref.~\cite{2101.08789}.
For concreteness we focus on massless axions coupled to two photons via the axion-photon coupling \gagg.
This coupling allows axions to be produced through the Primakoff process~\cite{10.1103/PhysRev.81.899}, where plasmons are converted into axions in the presence of electromagnetic fields generated by electrons or ions.

Primakoff production is the dominant axion production process in the Sun for $\ma \lesssim \si{meV}$ as long as the axion-electron coupling \gaee is sufficiently weak, $\gaee \ll \num{0.01}\,\gagg/\si{\GeV^{-1}}$~\cite{2101.08789}.
However, once $\gaee \sim \num{0.01}\,\gagg/\si{\GeV^{-1}}$, its additional contributions to the axion flux cannot be disregarded.
While this requires an extension of the methodology presented here, we do not foresee any fundamental obstacles to this (see \cref{sec:comments} for further comments).
It is important to note that an axion detection in a helioscope implies $\gagg \neq 0$.

\subsection{The expected helioscope signal from the Primakoff process}

The Primakoff production rate of axions in the Sun, including the effects of charge screening via the Debye screening scale~$\ks$~\cite{Raffelt:1985nk,Raffelt:1987np}, is given by 
\begin{equation}
	\GP(E_a) = \frac{\gagg^2 \, \ks^2 \, T}{32\pi} \left[\left(1+\frac{\ks^2}{4E_a^2}\right) \, \log\left(1+\frac{4E_a^2}{\ks^2}\right)-1\right] \frac{2}{\ee^{E_a/T}-1} \, , \label{eq:Primakoff_rate}
\end{equation}
where $T$ is the temperature inside the Sun and $E_a$ is the axion's energy.
The quantities \ks and $T$ are not entirely independent since \ks can be expressed as~\cite{Raffelt:1985nk}
\begin{equation}
    \ks^2 = \frac{4\pi \alphaEM}{T} \left(n_e + \sum_{z}Q_z^2 \, n_z \right) \, .
	\label{eq:ks}
\end{equation}
In the expression above, $\alphaEM \approx 1/137$ is the fine-structure constant, while $n_z$ and $Q_z$ are the number density and electric charge (in units of the elementary charge) of each ion species (labelled by $z$) in the plasma.

\Cref{eq:Primakoff_rate} holds for relativistic axions, i.e.\ $E_a \gg \ompl$.
Here, $\ompl$ is the plasma frequency, which is less than about \SI{0.3}{\keV} inside the Sun.
Moreover, \cref{eq:Primakoff_rate} neglects electron degeneracy effects, which reduce the Primakoff axion flux by about 3\% at most~\cite{2101.08789}.
Incorporating them would complicate our analysis, and we disregard them for simplicity.
One might expect that such small systematic shifts only mildly affect our fitting procedure and results.
However, as we will see in \cref{sec:results}, once a sufficiently high number of axions are detected, estimating $T$ and \ks from the approximate \cref{eq:Primakoff_rate} while calculating the event rates with the full result can lead to problems with the fitting procedure and systematic shifts in the inferred solar properties.
While this is only a small issue for the reconstruction of $T$, the weaker dependence of the Primakoff rate on \ks leads to much larger deviations.
In a more refined setup this could be addressed by allowing for a more realistic fit function that takes the electron degeneracy into account, cf.\ ref.~\cite{2101.08789}.

\begin{figure}
    \centering
    \includegraphics[width=4.5in]{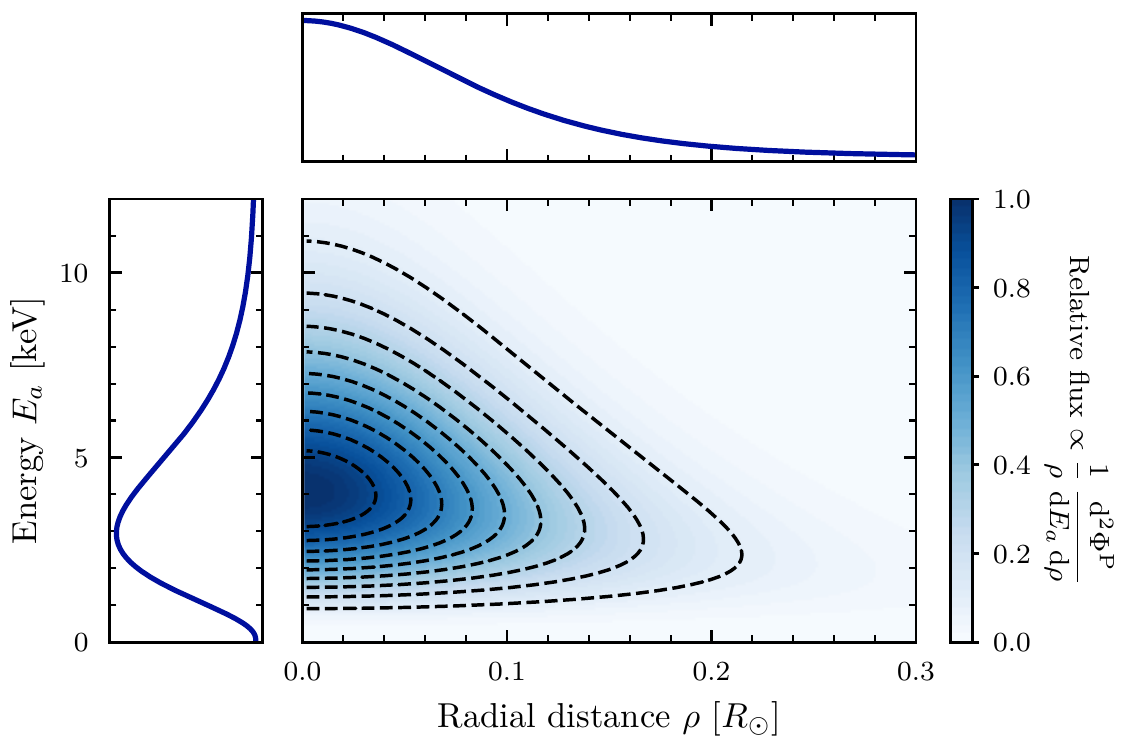}
    \caption{Distribution of the differential B16-AGSS09 solar axion flux $\frac{1}{\rho}\,\frac{\dd^2 \PhiP}{\dd E_a \, \dd \rho}$, normalised to its maximum value. The central panel shows the full distribution, while the adjacent panels show the relative marginal fluxes, i.e.\ where either $\rho$ or $E_a$ has been integrated out.}
    \label{fig:flux_distribution}
\end{figure}
Furthermore, we can ignore parallaxes due to the vast difference in scale between the solar radius and the distance between Earth and the Sun, $\Rsol \approx 0.005\,\dE$, as discussed in e.g.\ ref.~\cite[sec.~2.6]{2101.08789}.
Consequently, we can project the axion flux (in units of $\si{\cm^{-2} \s^{-1} \keV^{-1}}$) from the solar disc onto the helioscope detector surface on Earth.
The spatial and spectral differential axion flux at Earth is given by~\cite[e.g.][sec.~2.6]{2101.08789}
 \begin{equation}
	\frac{\dd^2 \PhiP}{\dd E_a \, \dd \rho} = \frac{\Rsol^3 E_a^2}{2 \pi^2 \dE^2} \, \int_{\rho}^{1} \! \dd r \; \frac{\rho \, r}{\sqrt{r^2 - \rho^2}} \; \GP(E_a, \, r) \, , \label{eq:axionflux}
\end{equation}
where $\rho$ is the distance from the centre of the solar disc in units of \Rsol.
\Cref{fig:flux_distribution} shows the resulting differential axion flux on the solar, which we compute numerically utilising the publicly available \code{SolarAxionFlux} library~\cite{2101.08789,solaraxionflux}.

The differential axion flux in \cref{eq:axionflux} is then converted into a differential photon flux inside the magnetic field of a helioscope.
The converted photons will have the same energy as the incoming axions, and we can set $\omega = E_a$.
The conversion probability after traversing a magnetic field of effective strength $B$ and length $L$ is given by
\updated{
\begin{equation}
    P_{a\gamma} = \frac{\omega}{\sqrt{\omega^2 - \ma^2}} \left(\frac{\gagg B L}{2}\right)^2 \mathrm{sinc}^2\left(\frac{q L}{2}\right) \rightarrow \frac14 \, \gagg^2 B^2 L^2 \quad (\ma \rightarrow 0) \, , \label{eq:pagamma}
\end{equation}
where $\mathrm{sinc}(x) \equiv \sin(x)/x$ and $q \equiv \omega - \sqrt{\omega^2 - \ma^2}$.
Conveniently, for light axions with masses $\ma \lesssim \SI{0.01}{\eV}$, the conversion probability $P_{a\gamma}$ is essentially independent of $\omega$.
}

To obtain the number of detected photons, we need to multiply \cref{eq:axionflux} by \cref{eq:pagamma}, the data-taking time $\Delta t$, and the effective exposure $\mtx{A}{eff} = \epsilon A$, which is the product of the physical detector cross section $A$ and the total efficiency $\epsilon$ of the X-ray optics and detector.
We also need to average the Primakoff rate over the relevant energy range.
For energies in the interval $\omega \in [\omega_j,\,\omega_{j+1}]$, we define the corresponding averaged rate $\GPB_j$ as
\begin{equation}
    \GPB_j(r) \equiv \int_{\om{j}}^{\om{j+1}} \! \dd \omega \; \frac{\omega^2}{2\pi^2} \,  \GP(\omega,\,r) \, . \label{eq:gpbar}
\end{equation}
With this definition, the expected number of photons in the $i$th radial and $j$th spectral bin, $\rho \in [\rh{i},\,\rh{i+1}]$ and $\omega \in [\omega_j,\,\omega_{j+1}]$, can be computed as
\begin{equation}
    \bar{n}_{i,j} = \frac{\Pag \, \mtx{A}{eff} \, \Delta t \, \Rsol^3}{\dE^2} \, \int_{\rh{i}}^{\rh{i+1}} \! \dd \rho \; \int_{\rho}^{1} \! \dd r \; \frac{r \, \rho}{\sqrt{r^2 - \rho^2}} \; \GPB_j(r) \, . \label{eq:expected_counts}
\end{equation}
Note that the $\bar{n}_{i,j}$ depend on the two functions of interest, $T(r)$ and $\ks(r)$.
However, due to the finite amount of data available, we can only hope to infer these functions at a finite number of points, \rd{i}.
These values correspond to points $\rh{i}$ on the solar disc, which we will identify with one another throughout our analysis.
In particular, this also implies that $\nr = \nrad$.

\subsection{Extraction of solar properties from data fitting}\label{sec:extraction}

In a helioscope, the photons resulting from axion conversion within the magnetic field have the same energy and follow the same direction as the incoming axions.
Using an energy-\ and position-resolving detector with sufficiently well characterised X-ray optics, a helioscope thus provides an ``axion image'' of the Sun i.e.\ the integrated axion flux emitted from the solar disc on the sky (cf.\ ref.~\cite[fig.~3]{hep-ex/0702006} or ref.~\cite[fig.~12]{10.1088/1475-7516/2015/12/008}).
Moreover, an axion false-colour image can be created based on the detected photon energies.

We now show that the axion image contains information about the different layers of the Sun and not just its bulk properties.
As can be seen from \cref{eq:axionflux}, the axion flux at radius $\rho$ on the detector is obtained from integrating over the solar radius $r$ along the direction perpendicular to the solar disc.
To reconstruct its original $r$~dependence from the dependence on the disc parameter $\rho$, we need to ``invert'' the integral relation \cref{eq:axionflux}.
This inversion enables the reconstruction of the underlying solar parameters, $T(r)$ and $\ks(r)$, at various points $\rd{i}$ within the Sun.

One possible approach involves interpolating the functions $T(r)$ and $\kappa_{s}(r)$ based on a chosen set of points \rd{i}.
However, only $\GPB_j(r)$ appears in \cref{eq:expected_counts}.
It is thus technically more straightforward and closer to the measured quantities to interpolate $\GPB_j(r)$ at the different \rd{i}.
The values for $\GPB_j(\rd{i})$ can be computed from $T_i \equiv T(\rd{i})$ and $\kappa_i\equiv \ks(\rd{i})$ using \cref{eq:gpbar}.

\updated{We interpolate the $\GPB_j(r)$ using splines.
This interpolation scheme can lead to ``ringing'', resulting in unphysical, negative values in the intervals between the \rd{i}.
This effect is most problematic for larger radii, where we expect a low number of photon counts.
Due to volume effects, even comparably small regions of negative $\GPB_j$ need to be compensated by e.g.\ systematically higher temperature values at smaller radii.
We prevent this issue by using piecewise-cubic Hermite interpolating polynomials (PCHIPs)~\cite{10.1137/0905021} (see \cref{app:poly_interp} for further discussion).}

Estimates for the true \gagg, $\kappa_i$, and $T_i$ from the helioscope data can then be inferred by fitting.
To this end, we use a Poisson-inspired fitting metric,
\begin{equation}
    \Delta \chi^2 \equiv -2 \log L(\gagg,\,\left\{\kappa_i,\,T_i\right\}) = 2 \sum_{j} \bar{n}_{i,j} - \hat{n}_{i,j} \, \log(\bar{n}_{i,j}) - \hat{n}_{i,j} + \hat{n}_{i,j} \, \log(\hat{n}_{i,j}) \, , \label{eq:fitting_metric}
\end{equation}
where $\hat{n}_{i,j}$ is an estimate for the number of photons $n_{i,j}$ observed in the $i$th spatial annulus and the $j$th spectral bin.\footnote{In general, the $\hat{n}_{i,j}$ are non-integer because we reconstruct a circle from a square detector.
As a consequence, the likelihood in \cref{eq:fitting_metric} could be based on, e.g., a mixed Gamma-Poisson process, whose parameters can be estimated from our error analysis shown in the right panel \cref{fig:detector_setup}.
While this would provide a more accurate description, especially in the regime of smaller counts, we argue that \cref{eq:fitting_metric} is sufficient for a proof of principle for our method.}
The expected number of counts $\bar{n}_{i,j}$ is given by \cref{eq:expected_counts}.

Since we need to determine $2\nr + 1 = 2\nrad + 1$ parameters in total, we expect that $\nerg \geq 3$ energy bins are needed, lest the fit is underdetermined.
We verified that choosing $\nerg = 1$ leads to the expected degeneracies between the $\kappa_i$ and $T_i$ coefficients, which start to disappear for $\nerg \geq 2$.

\updated{In practice we also need that the information in the different bins provides sufficient discriminating power to determine the $T_i$ and $\kappa_i$.\footnote{\updated{Note that we fit both quantities simultaneously. However, in this paragraph we disentangle the process into two steps to obtain a qualitative understanding.}}
For the temperature, this is not an issue: the total (energy integrated) number of relativistic axions produced via the Primakoff process scales with $T^6$~\cite[e.g.][sec.~5.2.1]{Raffelt:1996wa}.
As we will see, this strong dependence allows to infer the $T_i$ rather well.
To also infer $\kappa_i$, we need to take into account the information from the energy dependence.
Unfortunately, \ks only weakly depends on energy through the ratio $2E_a/\ks$, as can be seen from the Primakoff rate on \cref{eq:Primakoff_rate}.
Since $T > \ks$, we have $2E_a/\ks < 1$ for axion energies where the flux is high.
Expanding \cref{eq:Primakoff_rate} in $2E_a/\ks$, one can see hat this dependence is rather weak.
This will make the fitting of this quantity more problematic.
}

\section{Case study for the IAXO helioscope}\label{sec:case_study}

\begin{table}
    \caption{Parameter values used in our case study. \textit{Left:} Parameters for the for ``IAXO baseline'' setup~\cite[table~5]{1904.09155}, except that we assume a runtime for IAXO of \SI{6}{\year} i.e.\ \SI{3}{\year} of data. \textit{Right:} Detector grid and simulation-related parameters.}
    \label{tab:params}
    \centering
    {
    \hfill
    \begin{tabular}{@{}lS[table-format=2.2]@{\,}l@{}}
    \toprule
    IAXO parameter & \multicolumn{2}{l}{Value} \\
    \midrule
    Magnetic field $B$ & 2.5 & \si{\tesla} \\
    Length $L$ & 20.0 & \si{\metre} \\
    Cross section $A$ & 2.26 & \si{\metre^2} \\
    Total effective efficiency $\epsilon$ & \num{0.56} & \\
    Data-taking time $\Delta t$ & 3.0 & \si{\year} \\
    \bottomrule
    \end{tabular}
    \hfill
    \begin{tabular}{@{}lr@{}}
        \toprule
        Parameter & \multicolumn{1}{l}{Value} \\
        \midrule
        Grid bins per side $\npx$ & 128 \\
        Energy bins \nerg & 4 \\
        Radial bins $\nrad = \nr$ & 20 \\
        MC simulations $\mtx{N}{\tiny MC}$ & 1000 \\
        \bottomrule
        & \\
    \end{tabular}
    \hfill
    }
\end{table}
To illustrate our method, we investigate the expected reconstruction abilities in the upcoming solar helioscope IAXO~\cite{1401.3233,1904.09155,2010.12076}.
We provide the associated experimental parameters in \cref{tab:params}, noting that the radial bins in the range $r \in [0,\,1]\,\Rsol$ are equally spaced while the spectral bins $\omega \in [0.3,\,15.0]\,\si{\keV}$ are chosen to equally distribute the observed number of counts between them.
The two benchmark models that we consider are very light axions ($\ma = 0$) with couplings of $g_{10} = 0.6$ (\case{A}) and $g_{10} = 0.1$ (\case{B}), where $g_{10} \equiv \gagg/\SI{e-10}{\GeV^{-1}}$.
\case{A} corresponds to a detection below the CAST limit of $g_{10} < 0.66$~\cite{1705.02290}, while the amount of counts in \case{B} would typically still lead to an axion detection in excess of $5\sigma$, even when including the background rates assumed for the ``IAXO baseline'' setup~\cite[table~5]{1904.09155}.
The total number of expected photons are $\bar{n} = \sum_{i,j} \bar{n}_{i,j} \approx \num{330000}$ and $\bar{n} \approx \num{250}$ photons for cases A and B, respectively.

\subsection{Experimental setup}\label{sec:setup}

\begin{figure}
    \centering
    \hfill
    {
    \includegraphics[width=205pt,valign=t]{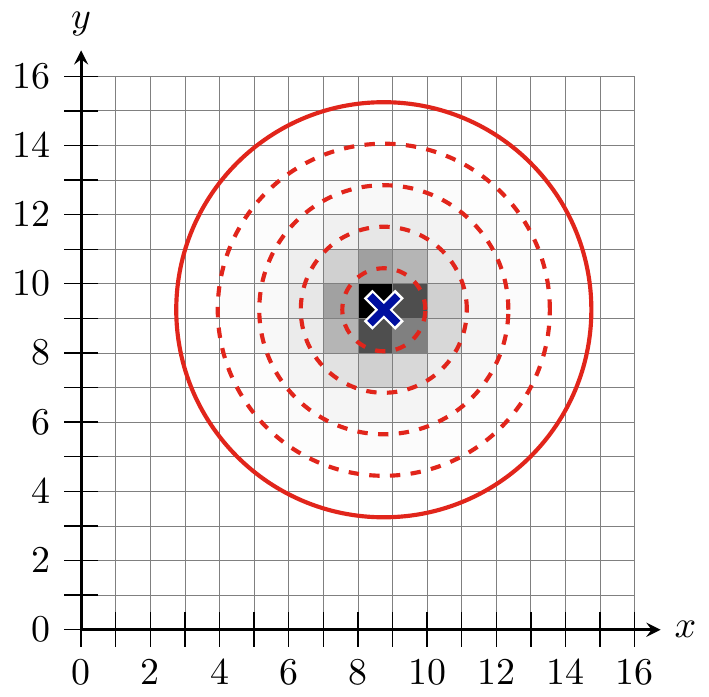}
    }
    \hfill
    \includegraphics[width=3in,valign=t]{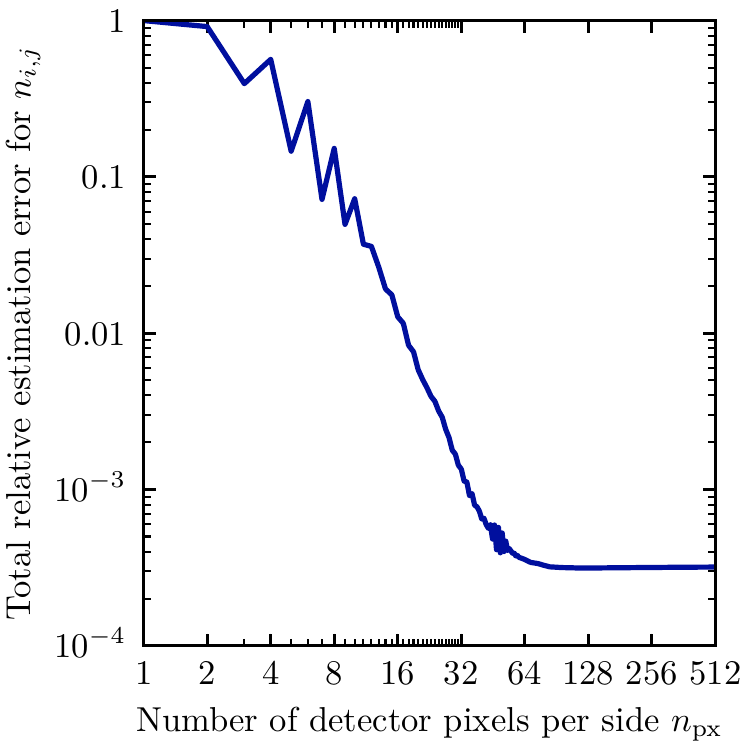}
    \hfill
    \caption{\textit{Left}: Axion-induced photon flux on the detector grid. The centre of signal region (blue cross) is at pixel coordinates $(8.75,\,9.25)$, where \Rsol corresponds to 6\,px. The solid red circle represents the boundary of the signal region, which is divided into five annuli (areas between the dashed red circles). The grey shading scales with the square root of the photon counts $n_{i,j}$, reflecting the magnitude of fluctuations. \textit{Right:} Total estimation error as a function of number of grid pixels per side, \npx.}
    \label{fig:detector_setup}
\end{figure}

The helioscope X-ray optics project the axion-induced photon signal onto an $\npx \times \npx$ detector grid, as shown schematically in the left panel of \cref{fig:detector_setup}.
The rings (dashed red lines) delimit the different annuli from which we infer the photon counts.
For illustrative purposes, we only show a grid of $16 \times 16$ photon-counting detector pixels.
However, in the subsequent analysis, we assume a grid of $128 \times 128$ pixels.
The proportions depicted in \cref{fig:detector_setup} correspond to our case study.

It is important to choose a sufficiently large value for \npx to accurately estimate the observed photon counts $n_{i,j}$.
Since the photon signal exhibits rotational symmetry, it makes sense to bin it radially in annuli.
However, it is mathematically impossible to exactly cover such radial bins with quadratic detector pixels (see \cref{sec:comments} for comments on ring-shaped detectors).
While the arising geometrical errors thus never vanish, they can be reduced by making the pixel size sufficiently small or, equivalently, \npx sufficiently large.

Furthermore, estimating the $n_{i,j}$ in each annulus requires an assumption about the spatial photon distribution in each pixel.
In our analysis we assume a uniform distribution, which we expect to be a good approximation as long as \npx is sufficiently large.
Still, as the signal increases towards the centre (cf.\ \cref{fig:flux_distribution}), a more quantitative investigation is warranted.
To address this, we examine the total error between the estimated $\hat{n}_{i,j}$ compared to the true $n_{i,j}$ as a function of \npx.
We employ the fitting metric in \cref{eq:fitting_metric} with the roles of $\hat{n}_{i,j}$ and $\bar{n}_{i,j} = n_{i,j}$ reversed, $g_{10} = 0.6$, and $\nerg = 1$, while keeping the relative proportions of the signal region shown in the left panel of \cref{fig:detector_setup}.
The right panel of \cref{fig:detector_setup} shows the total estimation error relative to its maximum.
As \npx increases, the error decreases while fluctuating, until it reaches the lowest possible values for $\npx \gtrsim 64$.
For larger values, the error appears to asymptotically settle at the lowest possible value.
This behaviour confirms our expectation that sufficiently large \npx should allow for an excellent estimate $\hat{n}_{i,j}$ of the photon counts contained in the annuli.
Although the exact shape of the estimation error may depend slightly on the relative location of the signal region, the size of the signal region's radius scales simply with the values of \npx.
Therefore, we can focus on the asymptotic region of sufficiently large \npx, where a relatively low error can be reliably guaranteed.

In addition to the considerations above, our setup relies on three additional assumptions.
First, that the X-ray optics do not distort or spread the axion image.
Second, that we have perfect control of the location and size of the axion image on the detector grid.
Third, that the background levels in IAXO are sufficiently low to justify ignoring detector dark counts or other backgrounds.
A detailed discussion of these assumptions is provided in \cref{sec:comments}.

\subsection{Solar model}\label{sec:solarmodel}

We choose the B16-AGSS09 solar model~\cite{0909.0948,1611.09867} for our case study.
\Cref{fig:flux_distribution} shows the associated two-dimensional differential solar axion flux, cf.\ \cref{eq:axionflux}, calculated using the publicly available \code{SolarAxionFlux} library~\cite{2101.08789,solaraxionflux}.
It is important to point out that the various standard solar models predict very similar $T(r)$ and $\ks(r)$.
For example, comparing the B16-AGSS09 and B16-GS98~\cite{10.1023/A:1005161325181,1611.09867} solar models, the respective $T(r)$ and $\ks(r)$ deviate less than 3\% across the whole Sun.
The most significant difference between the solar models appear in their metallicities, which we do not explicitly consider in this work.
We refer the reader to ref.~\cite{2101.08789}, where axions are used to distinguish low-\ and high-metallicity solar models.
In fact, the most recent generation of the solar models might resolve the metallicity problem altogether~\cite{2203.02255}.

\subsection{Fitting procedure}\label{sec:fitting}

Before describing the generation of IAXO pseudodata, let us comment on difficulties that can arise when optimising the fitting metric $\Delta \chi^2$ in \cref{eq:fitting_metric}.

One issue is that the flux is concentrated around the centre of the signal region on the detector grid, requiring a sufficiently large \npx (cf.\ \cref{sec:setup}).
We also need to choose radial bins to be somewhat evenly distributed across the entire range for a faithful reconstruction (cf.\ \cref{app:poly_interp}).
Since 99\% of all axions are produced within $r \lesssim 0.5\,\Rsol$, almost no data is available from beyond this.
The corresponding values of $T(r)$ and $\ks(r)$ can thus only be inferred indirectly from interpolating between the inner region and the edge of the Sun.
In this sense, the choice of interpolating function can have a noticeable impact of the results.
For example, choosing PCHIPs with $\nr = 5$ will introduce a 10\% error on $\bar{n}$, which only drops below 1\% for $\nr \gtrsim 20$.

\updated{As already discussed at the end of \cref{sec:extraction}, the dependence on $\ks$ is rather weak.}
In practice, this results in rather shallow minima of $\Delta \chi^2$ around $\kappa_i$, which presents a challenge for numerical optimisation.
This is in contrast to the strong dependence on the temperature $T$, which typically leads to steeper minima.
We use a combination of the adaptive Nelder-Mead algorithm~\cite{10.1093/comjnl/7.4.308,10.1007/s10589-010-9329-3} from \texttt{scipy}~\cite{2020SciPy-NMeth} and subsequently the MINUIT algorithm~\cite{10.1016/0010-4655(75)90039-9}, as implemented in \texttt{iminuit}~\cite{10.5281/zenodo.3949207}, to ensure convergence.\footnote{We checked that the optimisation algorithms can infer correctly the input parameters when the data corresponds to the expectation value of \cref{eq:expected_counts}, computed with the interpolated $\GPB_j(\rd{i})$. We also cross-checked our procedure with the Bayesian \texttt{emcee} sampler~\cite{1202.3665} and the \texttt{scipy} implementation of the heuristic global optimisation strategy of differential evolution~\cite{2020SciPy-NMeth,10.1023/A:1008202821328}.}

One may also consider expanding our fitting approach by incorporating additional likelihoods or incorporating physical prior information.
One option is to utilise more information from solar modelling (see ref.~\cite{2007.06488} for a review), allowing for a Bayesian analysis with priors informed by physical considerations.
For instance, we could leverage the knowledge that the ratio $\xi^2_i \equiv (\kappa_i/T_i)^2 \approx 48$ remains approximately constant within the relevant regions of the Sun, with an accuracy of about 15\%~\cite[p.~169]{Raffelt:1996wa}.
Additionally, data from helioseismology or neutrino experiments could be used.
For example, the Borexino experiment can determine $T(r \approx 0)$ to sub-percent precision~\cite{2205.15975}, thanks to the strong scaling of the \ce{^8B} decay flux with temperature~\cite[e.g.][]{0805.2013}.
Surface temperature measurements at $T(r \approx 1)$ could provide further insights.

We do not explore these possibilities in detail in this study, as our main focus is on presenting a model-independent reconstruction of $T$ and \ks.
However, it is worth noting that the established practice of ``global fits'' in solar modeling indicates the potential for complementary information to enhance our understanding after detecting axions.

\subsection{Monte Carlo simulations of IAXO pseudodata}\label{sec:mcsims}

The Monte Carlo (MC) procedure is used to generate photon pseudodata sets for IAXO.
This procedure involves the following steps, which are repeated $\mtx{N}{\tiny MC}$ times:
\begin{enumerate}[label=(\roman*),itemsep=0.5ex]
    \item Draw a random number of observed photons $n$ from a Poisson distribution, $n \sim \mathcal{P}(\bar{n})$, where the expected number of photon counts $\bar{n}$ is given by \cref{eq:expected_counts}.\label{step:draw}

    \item Determine the \nerg energy bin boundaries such that each bin contains approximately the same number of photon counts.\label{step:split}
    
    \item Distribute the $n$ photons on \nerg spectral detector grids according to the spectral and radial differential flux (cf.\ \cref{fig:flux_distribution}) and rotational symmetry.\label{step:grid}

    \item Estimate the photon counts $\hat{n}_{i,j}$ for each of the \nrad annuli on the \nerg grids.\label{step:estimate}

    \item Fit \gagg and the coefficients $\kappa_i$ and $T_i$, as described in \cref{sec:fitting}.\label{step:fit}
\end{enumerate}
The 20 radial bins are distributed equally in the interval $r \in [0,\,1]$.
All other quantities are set to the values given in \cref{tab:params}.

\subsection{Expected reconstruction abilities for solar properties}\label{sec:results}

\begin{figure}
    \centering
    \includegraphics[width=6in]{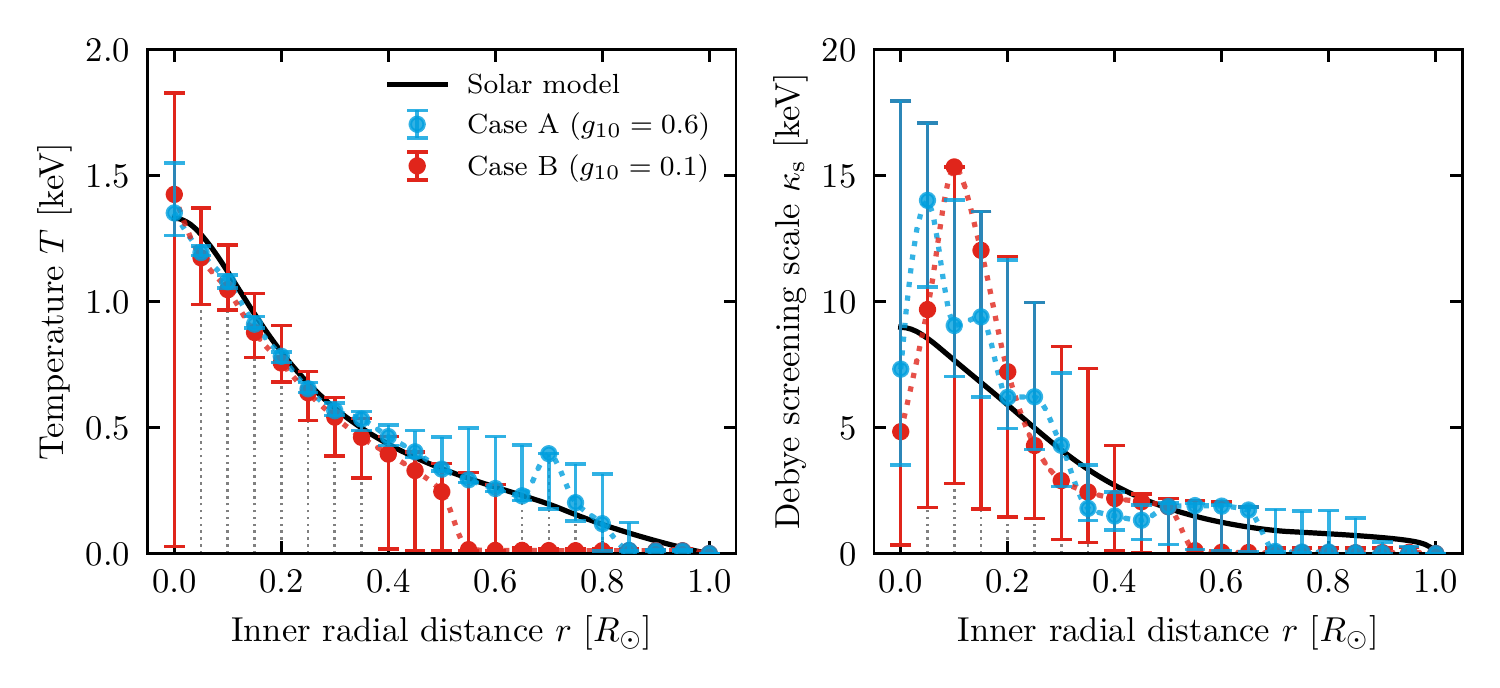}
    \caption{Results for the inferred radial profile of $T$ (\textit{left}) and \ks (\textit{right}). Black lines indicate true values from the B16-AGSS09 solar model. Blue and red points and error bars show the median and 68\% central intervals best-fitting parameter values for \case{A} and \case{B}, respectively. The dotted lines between the median values are only to guide the eye as we do not assume any interpolating function for $T$ and \ks.}
    \label{fig:results}
\end{figure}
\Cref{fig:results} shows the results of our case studies.
Comparing to the reference solar model (black lines), we immediately see that the temperature profiles can be determined fairly accurately up to about $0.8\,\Rsol$ in \case{A} and $0.4\,\Rsol$ in \case{B}.
The median accuracy in these regions, as computed by the absolute deviations from the solar model values, is about 3\% and 7\% for the two cases, respectively.
Of course, we will only obtain one IAXO data set over the course of our measurement period, and we thus need to also consider the precision with which we expect to typically determine the $T_i$ in the relevant regions.
We measure it by determining the median precision within the relevant region, where the individual precision of the $T_i$ are computed as half the size of the central 68\% interval of all best-fitting points compared to the respective median values.
Doing so, we find a precision of better than 10\% in \case{A} and 16\% in \case{B}.

These findings are not unexpected, given that 99\% of the photon counts can be found within $\rho \lesssim 0.5\,\Rsol$, as mentioned previously.
While the continuity of the {\GPB}s and assumptions about the interpolating function helps to infer information for larger $\rho$, and thus $r$, there is still a fundamental limit given by the number of counts.
This is also why the best-fitting points for larger radii in \cref{fig:reconstruction} can be found at values close to zero.

Continuing the discussion with observed systematic deviations, we observed that these mostly stem from the simplified Primakoff rate in \cref{eq:Primakoff_rate}, but also the residual estimation and interpolation errors from having to integrate over the square detector grid and our choice of $\nrad = 20$.
By generating data based on the interpolated, rather than the solar model values of the {\GPB}s, we can verify that the systematics mostly disappear.
This also allows us to exclude major issues in the minimisation procedure, which however still adds a small contribution due to the relatively high number of parameters to be optimised.
In this sense, we expect there to be a trade-off between reducing systematic errors by further increasing \nrad, and still maintaining a reliable fitting procedure given a finite amount of computational resources.

Regarding the observed precision, it is clear that the indirect reconstruction procedure with a non-trivial algorithm behind the PCHIPs does not allow for a straightforward error propagation for the {\GPB}s compared to the alternative interpolations discussed in \cref{app:reconstruction}.
However, the errors on $T$ and $\ks$ do not seem to follow the simple a simple scaling relation based on the ratio of \gagg in \case{A} and \case{B} either.
In particular, we also observe a balancing of opposing effects for achievable precision: more photon counts towards the central region of the detector (smaller uncertainty) balance against accumulating errors towards the centre, cf.\ \cref{eq:pwc_errors}, and a volume effect for our equally spaced \rd{i} (larger uncertainties).

Considering the radial profile of $\ks$, we find a median accuracy of about 30\% for both cases in the regions where we can determine $T(r)$ with good accuracy.
Conversely, the median precision is about 50\% for \case{A} and 90\% for \case{B}.
This means that a model-independent determination within our method becomes challenging for \ks.
As explained before, \updated{this is due to the shallow minima of $\Delta \chi^2$ (maxima of $L$) with respect to the $\kappa_i$.}

To guide the eye, we connect the median best-fitting $T_i$ and $\kappa_i$ values with PCHIPs i.e.\ the interpolating splines that we assumed for the {\GPB}s (see \cref{app:poly_interp}).
However, we did not make any assumptions on the shape of $T(r)$ and $\ks(r)$, and our analysis can thus only determine the $T_i$ and $\kappa_i$ at the pre-selected values of $r_i$.
Even for a piecewise-constant interpolation of {\GPB}, discussed in \cref{app:constant_interp}, the underlying $T(r)$ and $\ks(r)$ could in principle be very complicated functions, as long as the resulting {\GPB}s are piecewise constant.

Overall, our method can yield a fairly precise estimate for $T(r)$ in at least the inner half of the Sun, even with a moderate number of detected photon counts as in \case{B}.
The small inaccuracies in the reconstruction could be reduced by using a more accurate Primakoff rate for the modelling and more detector pixels.

\section{Discussion of limitations and related methods}\label{sec:comments}

This section is to summarise our simplifying assumptions and to discuss problems and their possible solutions in more realistic settings.
We also comment on the applicability of our results to neutrino experiments.

\subsection{Limitations and extensions}

\paragraph{More general axion models.}
We assumed effectively massless axion models that predominately couple to photons.
While helioscopes typically cannot distinguish axion models with $\ma \lesssim \si{\meV}$ from the truly massless case, it is known that IAXO can determine an axion mass of $\ma \gtrsim \si{\meV}$ at the $3\sigma$ level~\cite{1811.09290}.
Similar to \gagg, \ma is a global parameter, which thus affects all contributions to $\Delta\chi^2$.
For this reason, we do not expect that the inclusion of \ma presents a major difficulty for the fitting procedure.
However, the computation will become more involved since the conversion probability in \cref{eq:pagamma} now depends on $E_a$, whose dependence cannot be factored out anymore.

When considering additional axion couplings such as \gaee, more computational complications arise.
This is because the associated rates now depend on many more solar parameters -- in principle (combinations of) all solar abundances!
In theory, this can be accounted for by additional spectral information, i.e.\ by increasing \nerg.
This will necessarily dilute the total number of photons per energy bin, which consequently introduces a larger level of uncertainty for all fitted parameters.
Due to this and the large number of fitting parameters, we expect an extension to \gaee to be possible but challenging in practice.

\paragraph{Helioscope detectors.}
As demonstrated in \cref{fig:detector_setup}, the accuracy of our inference of the $\GPB_j$s relies on having sufficiently large \npx.
This compensates for the systematic error introduced by our assumption, used in \cref{step:estimate}, that the fractional photon counts are equally distributed in each pixel.
Helioscope detectors with enough pixels exist, such as the grid-based pn-CCD detector used in CAST with $200 \times 64$ pixels \cite{physics/0702188}.
Gridpix detectors with $\npx = 256$ have also been used successfully~\cite{10.1016/j.nima.2017.04.007}.

We also ignore intrinsic and extrinsic backgrounds, as their rates in the context of IAXO are assumed to be very low and close to the background-free case~\cite[e.g.][]{1904.09155}.
For instance, CAST achieved integrated background rates of \SI{4.44e-5}{\per\s} in the \SIrange{1}{7}{\keV} energy range in 2004~\cite{physics/0702188}.
This corresponds to a total of about \num{1400} photons for $\Delta t = \SI{3}{\year}$, which is larger than the number of signal photons expected in \case{B}.
However, we anticipate improved detector capabilities for the next generation of axion helioscopes.
Additionally, the background rate can be determined extremely accurately in helioscopes before sunrise and after sundown.
This calibration allows for background subtraction or modelling in the statistical inference step.

\updated{
Other detector types, such as Micromegas detectors~\cite{10.1016/0168-9002(96)00175-1} or metallic magnetic calorimeters (MMCs; see refs~\cite{10.1063/1.3292407,10.1109/TASC.2009.2012724,10.1007/s10909-018-1891-6} for reviews) could also be well-suited for IAXO.
MMCs have different systematics than gaseous detectors and offer excellent energy resolution, making them an attractive additional technology~\cite[e.g.][]{2010.15348}.
As previously mentioned, this could enable IAXO to test axion models and solar properties after a discovery.
}

Scaling up an available MMC prototype~\cite{2010.15348} in a {na\"ive} manner, it seems feasible to achieve $\npx = 48$ with background rates of \SI{0.002}{\per\s} in the \SIrange{1}{10}{\keV} energy range.
Notably, this background rate does not consider potential reductions through active muon vetos~\cite{GastaldoPrivateComm}, which have demonstrated up to a factor of 2 reduction~\cite{1312.4282}.
Additionally, the production of MMCs with ``ring-like'' pixels~\cite{GastaldoPrivateComm} appears feasible, potentially eliminating the need for the reconstruction step in estimating $n_{i,j}$.
While this would remove the associated systematic uncertainties, the location and shape of the signal region must be well-calibrated, stable, and accurately projected during operation to benefit from the technological advantages.
While promising, MMCs will require further development for improved detector properties and careful consideration of the challenges associated with maintaining experimental stability when utilising a detector with rotational symmetry.

\updated{
Micromegas detectors at present offer even lower background rates, currently demonstrating more than two orders of magnitude reduction compared to the previously mentioned (unshielded) MMC rate~\cite{10.1016/j.nima.2022.167913}.
This brings us closer to our assumption and the IAXO target of an effectively background-free experiment.
Additionally, Micromegas detectors may simplify the analysis procedure: when photons interact with the Micromegas gas volume, they trigger electrons to traverse an electric field.
The drift electrons are then amplified and recorded on detector strips, allowing us to infer the initial photon hit position from the strip signals.
For strips with a pitch of $\Delta x$, i.e.\ the distance between the centres of adjacent strips, the position uncertainty around a single strip is given by $\sigma_x = \Delta x / \sqrt{12}$~\cite[ch.~2]{10.1017/CBO9780511534966}, and even lower when hitting multiple strips.
Assuming that the axion image of the Sun fills most of the detector surface (cf.\ \cref{fig:detector_setup}), we can convert the position uncertainty into a number of effective equivalent pixels to directly compare the two cases.
Given a detector size of $s = \SI{60}{\mm}$ and pitch $\Delta x = \SI{0.5}{\mm}$~\cite{10.1016/j.nima.2022.167913}, the Micromegas detector would be equivalent to a Gridpix-like detector with $\npx \sim s/\sigma_x \approx 416$.
Therefore, even accounting for positional uncertainty, Micromegas detectors can simplify our procedure by eliminating the need for a pixel-based photon detector and directly using inferred photon hit positions.
}

\paragraph{X-ray optics and calibration.} 
Accurate tracking of the Sun and faithful projection of the axion image are critical to the success of the helioscope experiment as well as our methodology.
The CAST experiment has already demonstrated a pointing accuracy of ``well below 10\% of the solar radius''~\cite{1705.02290}.
Compared to the size of our radial bins, uncertainties of order $0.1\,\Rsol$ would have to be included via a point-spread function~(PSF), so a more precise determination of the pointing accuracy is needed.
After an axion detection, the centre and accuracy of the signal region can be cross-checked by averaging the position of all detected photon counts on the detector grid and computing their standard deviation, respectively.
The expected size of the ``axion image'' can be calibrated using previously established techniques~\cite[fig.~3]{hep-ex/0702006}.

Another issue concerns imperfect optics, which can lead to an overall reduction in the number of detected photons and spectral distortions, which violate the simplistic assumptions made in this work.
Nevertheless, if the effects of imperfect optics can be calibrated and described mathematically, they present a technical rather than a fundamental problem.
In fact, our setup already includes signal reduction due to imperfect optics.

The finite energy resolution of the detectors can be taken into account by convolving the expected signal with a smoothing kernel for a given energy resolution.
The energy resolution for the Gridpix detectors has been measured to be around 10--20\%~\cite{1709.07631}.
Given that the size of the energy bins used in this experiment is at least \SI{1.2}{\keV}, the effect of finite energy resolution is expected to be small.

It is also important to note that detectors have a finite energy threshold.
The lower end of the energy range used in this experiment, $\omega \geq \SI{0.3}{\keV}$, is reasonable but realistically achievable thresholds might be slightly higher~\cite{GastaldoPrivateComm}.

Regarding distortion and shear from X-ray optics, it is possible to include its effects in the analysis as long as the shape of the PSF can be determined.
However, the computations become more challenging if the PSF is highly asymmetric or involves shear, as suggested by e.g.\ ref.~\cite[fig.~3]{1705.02290}.
In this case, fitting the expected counts in each grid bin directly may be more appropriate instead of approximating the counts in the annuli.
However, this approach is computationally very challenging and requires further investigation to improve its feasibility.

\subsection{Connections to helioseismology and neutrinos}

\paragraph{Inversion techniques in helioseismology.}
Inversion techniques, such as the one developed in this work, have been studied in many different contexts in the past.
In solar physics, we should highlight techniques for inferring solar properties from the observation of helioseismic activity on the solar disk (see e.g.\ refs~\cite{10.1098/rsta.1984.0080,10.1038/310022a0} for early works).
Starting from ref.~\cite{10.1038/315378a0}, using observations from ref.~\cite{10.1038/302024a0}, the sound speed profile $c^2(r)$ throughout the Sun could be reconstructed in a model-independent way i.e.\ allowing to compare to the predictions of solar models.
This is also true for the solar density profile $\rho(r)$ (see
ref.~\cite{10.1007/BF00158422} for an early review of inversion techniques).

Making further assumptions about the energy transport in the solar core, helioseismic data can also be used to constrain the solar temperature using a polynomial ansatz~\cite{10.1086/175451}, which is similar to our approach.
This becomes possible thanks to a relationship of the sound speed and the ratio $T(r)/\mu(r)$, where $\mu$ is the mean molecular weight.
The uncertainty of the inferred temperature profile is about 3\%.
Thus, at the cost of arguably mild, additional assumptions and some model dependence, helioseismology allows for an accurate determination of $T(r)$ in at least the innermost region of the Sun.

\paragraph{What about neutrinos?}
Given the parallels between axion and neutrino phenomenology, it is natural to ask if we could achieve the same results with the already available neutrino data.
Indeed, it has long been recognised that neutrinos provide insights into the \emph{central} temperature of the Sun~\cite[e.g.][]{hep-ph/9309292,astro-ph/9401024}, which led to them being referred to as ``solar thermometers'' as well~\cite{hep-ph/9408277}.
Through the observed neutrino flux, we can measure an effective temperature $\mtx{T}{eff}$ of the Sun, which involves an integration over the whole solar volume.
As for axions, $\mtx{T}{eff}$ is close to the core temperature of the Sun due to the strong temperature dependence of neutrino interactions.
Still, this is not the same as determining the radial dependence of $T(r)$, for which we need to use a spatially-resolved ``neutrino image'' of the Sun.

While neutrino observatories can generate a solar neutrino image,\footnote{See the famous ``neutrino image'' from the Super-Kamiokande (Super-K) detector, e.g.\ available at \url{https://apod.nasa.gov/apod/ap980605.html}, or an updated version on their website \url{https://www-sk.icrr.u-tokyo.ac.jp/en/sk/about/research/}).} the effective PSF of the experiments blurs them significantly.
The angular resolution of these experiments is indeed quite poor as, e.g., shown in refs~\cite[figs~17, 19, 20]{1606.07538}, \cite[fig.~40]{hep-ex/0508053}, or \cite[fig.~26]{1010.0118}.\footnote{Note that the Sun is in the direction of $\cos(\theta) = 1$ in these figures.}
This is unfortunate given that, over its entire lifetime, the Super-K experiment has observed more than \num{e5} neutrinos~\cite{2202.12421}, which is comparable to the number of axions considered in \case{A}.
Recent results from the Borexino experiment, who report about \num{e4} detected solar neutrinos, only manage to correlate the origin of the neutrinos with the general position of the Sun, rather than the specific location of neutrinos on the solar disc~\cite{2109.04770,2112.11816}.
However, the future Hyper-K experiment may have better resolution for solar neutrinos~\cite{1805.04163}.

\updated{
We also note that while current solar neutrino experiments operate in the \si{\MeV} range, future facilities might also detect the complementary \si{\keV} neutrino flux~\cite{nucl-th/0006055,1708.02248}.
}

In summary, while combining a future axion detection with neutrino measurements is an exciting prospect, it seems that axions would currently be superior to neutrinos in terms of mapping out the temperature dependence inside the Sun, thanks to the available X-ray technology.

\section{Summary and concluding remarks}\label{sec:conclusions}

We propose a new method to infer radial profiles of the solar temperature $T(r)$ and Debye scale $\ks(r)$ following the detection of axions in a helioscope.
The strong dependence of the axion production rate on $T$ enables a precise reconstruction of this quantity in IAXO after an axion detection.
The reconstruction of $\ks(r)$ is more challenging, and care should be taken to use accurate and consistent signal modelling when a large number of axions is detected.

Our method complements other probes of the solar interior such as solar sound profile measurements in helioseismology or neutrino experiments.
Axions and neutrinos provide immediate information since they reach us within a relatively short time scale, namely less than about \num{8.5} minutes after their creation.
They leave the solar core mostly unimpeded due to their small interaction rates, and their time resolution is only limited by the need to collect enough statistics.
In contrast, photons produced in the Sun's core take at least \num{100000} years to reach the surface~\cite{10.1086/172103} and are subject to a huge number of scatterings during that time.

The detection of photons resulting from axion conversion benefits from the availability of high-quality X-ray optics, currently offering superior spatial resolution compared to neutrino probes.
Our methodology allows for a model-independent reconstruction of solar temperature at various locations within the Sun, providing an edge over helioseismology, which relies on additional assumptions about the solar composition.

In addition to the forward modelling approach presented in the main text, we also derived an analytical reconstruction algorithm for piecewise-constant and regular spline interpolations of the energy-averaged Primakoff rates (see \cref{app:reconstruction}).
These algorithms could be valuable for studies aiming to solve similar inverse problems.
Alternatively, simulation-based inference techniques could be explored as an alternative~\cite[e.g.][]{1911.01429}.

Our procedure can be further enhanced by integrating complementary information from helioseismology, neutrino physics, or by applying physically-informed priors in a Bayesian framework, as discussed in \cref{sec:fitting}.
This would be particularly useful for investigating axion models with additional couplings or for addressing technical challenges in realistic experimental setups.

To facilitate future modifications and utilisation of our method, we provide the computational routines used in this study as Python scripts in an updated version of our \code{SolarAxionFlux} library on Github~\cite{solaraxionflux}.

\acknowledgments
{
\footnotesize
We acknowledge helpful discussions with Esther Ferrer-Ribas, Loredana Gastaldo, Igor Irastorza, Sebastian Schmidt, and Julia Vogel about the capabilities of IAXO detectors and optics, and with Thomas Schwetz about neutrino experiments.
We are also indebted to Alex Geringer-Sameth for his brilliant idea to simplify the integration of radially symmetric functions on a square grid.
SH has received funding from the European Union's Horizon Europe research and innovation programme under the Marie Sk{\l}odowska-Curie grant agreement No~101065579.
JJ would like to acknowledge support by the European Union's Horizon 2020 research and innovation programme under the Marie Sk{\l}odowska-Curie grant agreement No~860881-HIDDeN.
LT was funded by the \textit{Graduiertenkolleg} ``Particle physics beyond the Standard Model''~(GRK 1940).
This work was performed using the Cambridge Service for Data Driven Discovery (CSD3), part of which is operated by the University of Cambridge Research Computing on behalf of the STFC DiRAC HPC Facility (\url{http://www.dirac.ac.uk/}).
The DiRAC component of CSD3 was funded by BEIS capital funding via STFC capital grants ST/P002307/1 and ST/R002452/1 and STFC operations grant ST/R00689X/1.
DiRAC is part of the National e-Infrastructure.
We acknowledge the CINECA award under the ISCRA initiative, for the availability of high-performance computing resources and support.
We made use of the \code{BibCom} tool~\cite{bibcom}.
}

\appendix

\section{Reconstructing the averaged Primakoff rates}\label{app:reconstruction}

Here we discuss how to explicitly reconstruct the energy-averaged Primakoff rates $\GPB_j$ using (the estimated) helioscope photon count data $\hat{n}_{i,j}$.
For the reasons given in \cref{app:poly_interp}, we only infer the {\GPB}s indirectly in our fitting approach in the main text.
However, we deem it instructive to present two pedagogical examples (piece-wise constant and regular spline interpolation schemes) for an explicit reconstruction of the {\GPB}s in \cref{app:constant_interp,app:spline_interp}.
This also provides the background for the procedure used in the main text, which we describe in \cref{app:poly_interp}.

Assume that the photon count data is binned in \nerg energy bins, labelled $\omega_j$, and in \nrad radial bins on the detector, labelled $\rho_i$ and in units of the solar radius.
In particular, the radial bins are sorted in increasing order, such that $\rh{1} = 0$ and $\rh{\nrad+1} = 1$.
The computation of the predicted number of photon counts $\bar{n}_{i,j}$ is given by \cref{eq:expected_counts}.

\subsection{Piecewise-constant interpolation}\label{app:constant_interp}

As a first approximation, consider a piecewise-constant interpolation for the $\GPB_j$:
\begin{align}
    \GPB_j(r) &= \sum_{i} \gamma_{i,j} \, \Theta(r - \rd{i}) \, \Theta(\rd{i+1} - r) \label{eq:Primakoff_ratePiecewise} \\
    \text{with} \quad \gamma_{i,j} &\equiv \int_{\om{j}}^{\om{j+1}} \! \dd \omega \; \frac{\omega^2}{2\pi^2} \,  \GP_i(\omega) \label{eq:gammaij} \\
    \text{and} \quad \GP_i(\omega) &\equiv \frac{\gagg^2 \, \kappa_i^2 \, T_i}{32\pi} \left[\left(1+\frac{\kappa_i^2}{4\omega^2}\right)\ln\left(1+\frac{4\omega^2}{\kappa_i^2}\right)-1\right] \frac{2}{\ee^{\omega/T_i}-1}  \, .
\end{align}
These equations depend on $\nerg\cdot\nrad$ constants $\gamma_{i,j}$, which we need to reconstruct in order to infer values for \gagg, $\kappa_i$, and $T_i$.

\subsubsection*{Matrix formalism}

We can use the ansatz in \cref{eq:Primakoff_ratePiecewise} to evaluate the integral in \cref{eq:expected_counts}, finding that
\begin{align}
    \bar{n}_{i,j} &\propto\int_{\rd{i}}^{\rd{i+1}} \! \dd \rho \, \rho \, \sum_{k=1}^{\nrad} \int_{\rho}^{1} \! \dd r \, \frac{r}{\sqrt{r^2 - \rho^2}} \, \gamma_{k,j} \; \Theta(r - \rd{k}) \, \Theta(\rd{k+1} - r) \\
    &= \int_{\rd{i}}^{\rd{i+1}} \! \dd \rho \, \rho \, \left[ \gamma_{i,j} \, \sqrt{\rd{i+1}^2 - \rho^2} + \sum_{k=i+1}^{\nrad} \gamma_{k,j} \, \left(\sqrt{\rd{i+1}^2 - \rho^2} - \sqrt{\rd{k}^2 - \rho^2} \right) \right] \\
    &= \frac13 \left[ \gamma_{i,j} \, \Delta_{i+1;i}^3 + \sum_{k=i+1}^{\nrad} \gamma_{k,j} \left( \Delta_{k+1;i}^3 - \Delta_{k+1;i+1}^3 + \Delta_{k;i+1}^3 - \Delta_{k;i}^3 \right) \right] \, , \label{eq:pwc_counts}
\end{align}
where we defined $\Delta_{m;n}^3 \equiv  (\rd{m}^2 - \rd{n}^2)^{3/2}$.
Note that \cref{eq:pwc_counts} can be written as a matrix equation of the form
\begin{equation}
    \bar{n}_{i,j} = \sum_{k=1}^{\nrad} \mathcal{M}_{ik} \, \gamma_{k,j} \, ,
\end{equation}
where the matrix $\mathcal{M}$ is defined as 
\begin{align}
    \mathcal{M}_{ik} = \frac{\Pag \, \mtx{A}{eff} \, \Delta t \, \Rsol^3}{3 \, \dE^2}
    \left\{\begin{array}{ll}
        \Delta_{i+1;i}^3 & \text{\quad for $i = k$,} \\
        \Delta_{k+1;i}^3 - \Delta_{k+1;i+1}^3 + \Delta_{k;i+1}^3 - \Delta_{k;i}^3 & \text{\quad for $k>i$,} \\
        0 & \text{\quad otherwise.}
        \end{array}\right.
\end{align}

\subsubsection*{Analytical solution of the matrix equation and error propagation}

Since $\mathcal{M}$ is an upper triangular matrix, it is straightforward to invert it, solving the underlying system of linear equations in the process.
We set $\gamma_{i,j} = 0$ for $i = \nrad + 1$, as it is at the edge of the Sun.
Then, the coefficients $\gamma_{i,j}$ for $i < \nrad + 1$ can be obtained in an iterative fashion by equating expected and observed counts, $\bar{n}_{i,j} = n_{i,j}$.
Note that this is a non-trivial approximation, which is why we need to simulate the behaviour of our method when applied to data.
This is further indicated by the fact that we can only estimate the true observed $n_{i,j}$, with non-integer estimators $\hat{n}_{i,j}$.
Keeping this in mind, we find that
\begin{align}
    n_{i,j} & = \sum_{k=i}^{\nrad}\mathcal{M}_{ik} \, \gamma_{k,j} =\mathcal{M}_{ii} \gamma_{i,j} + \sum_{k=i+1}^{\nrad} \mathcal{M}_{ik} \, \gamma_{k,j} \\
    \Rightarrow \quad \gamma_{i,j} &= \frac{1}{\mathcal{M}_{ii}} \left( \hat{n}_{i,j} - \sum_{k=i+1}^{\nrad} \mathcal{M}_{ik} \, \gamma_{k,j} \right) \, . \label{eq:pwc_solution}
\end{align}
This formula allows us to propagate the (independent) statistical errors from each radial bin for the reconstructed $\gamma_{i,j}$.
Assuming Gaussian error propagation, and approximate Poissonian errors for the photon counts, we find that
\begin{align}
    \sigma_{i,j}^2 \equiv \left(\Delta\gamma_{i,j}\right)^2 &= \frac{1}{\mathcal{M}_{ii}^2} \left[ \left(\Delta \hat{n}_{i,j}\right)^2 + \sum_{k=i+1}^{\nrad} \mathcal{M}_{ik}^2 \, \sigma_{k,j}^2 \right] \nonumber \\
    &= \frac{1}{\mathcal{M}_{ii}^2} \left[ \hat{n}_{i,j} + \sum_{k=i+1}^{\nrad} \mathcal{M}_{ik}^2 \, \sigma_{k,j}^2 \right] \, ,
    \label{eq:pwc_errors}
\end{align}
keeping in mind that $\sigma_{\nrad + 1,j}^2 = 0$.

The analytical expression in \cref{eq:pwc_errors} reveals that, due to the reconstruction procedure, the uncertainty accumulates towards the centre of the Sun, which is the primary region of interest.
This effect becomes particularly relevant when the number of observed photons is low.
Also note that, if we use roughly equally-spaced radial bins, the number of photons in the innermost bin is somewhat smaller due to the smaller volume.
This, too, increases the error for this bin.

\begin{figure}
    \centering
    \includegraphics[width=6in]{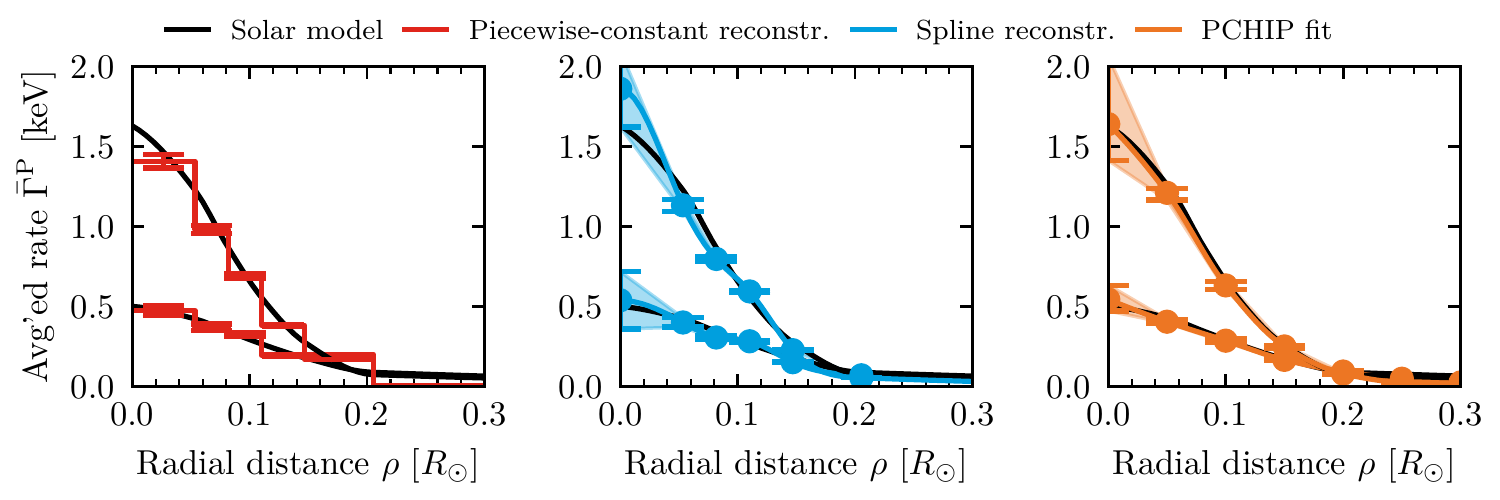}
    \caption{Reconstruction for the energy-averaged Primakoff rate \GPB (black line) for the lowest and highest of $\nerg = 4$ energy bins using piecewise-constant (\textit{left}), spline (\textit{middle}), and PCHIP (\textit{right}) interpolations. Error bars/shaded regions correspond to the standard deviation (68\% central region; \textit{right}). We use $g_{10} = 0.6$, the B16-AGSS09 solar model, and $\nrad = 6$ ($\nrad = 20$; \textit{right}) radial bins.}
    \label{fig:reconstruction}
\end{figure}
Reconstructions of the energy-averaged Primakoff rate \GPB are shown for three energy bins ($\nerg = 3$): piecewise-constant (\textit{left}), spline interpolation (\textit{middle}), and fitted PCHIP spline (\textit{right}).
Error bars represent the standard deviation, while shaded regions indicate the 68\% central region.
The solar model is B16-AGSS09 with $g_{10} = 0.6$, and $\nrad = 6$ ($\nrad = 20$) radial bins are used, along with $\nerg = 4$ spectral bins, in the first two cases (the PCHIP case).

The propagated uncertainties in \cref{eq:pwc_errors}, which we show in the left panel of \cref{fig:reconstruction}, can be taken into account when fitting for the coefficients \gagg, $\kappa_i$, and $T_i$.
Given the reconstructed coefficients $\gamma_{i,j}$, we can do so by optimising the following fitting metric:
\begin{equation}
    \Delta \chi^2 \equiv -2 \, \log L(\gagg,\,\left\{\kappa_i,\,T_i\right\}) = \sum_{j} \vc{x}_j^\mathrm{T} \, \Sigma^{-1}_j \, \vc{x}_j \, , \label{eq:reconstr_fitting_metric}
\end{equation}
where the entries of the vector $\vc{x}_j$ are given by $x_{i,j} \equiv \GPB_j(r_i\,|\,\gagg,\,\kappa_i,\,T_i) - \gamma_{i,j}$, and the covariance matrices $\Sigma_j = \mathrm{diag}(\sigma_{1,j}^2,\, \dots,\, \sigma_{\nrad,j}^2)$ are given by \cref{eq:pwc_errors}.

\subsection{Standard cubic spline interpolation}\label{app:spline_interp}

The piecewise-constant ansatz in \cref{eq:Primakoff_ratePiecewise} is only one possible option.
Cubic spline interpolation could be a more realistic choice which, however, introduces additional coefficients and requires the choice of boundary conditions.
Taking into account that the function values, and their first and second derivatives, need to be continuous across radial bin boundaries, the most general cubic spline ansatz can be written as~\cite[sec.~3.3]{numerical_recipes}

\begin{align}
    \GPB_j(r) = \sum_i &\Bigg[ \frac{\mu_{i,j}}{6 h_i} \, \left(\rd{i+1} - r\right)^3 + \frac{\mu_{i+1,j}}{6 h_i} \, \left(r - \rd{i}\right)^3 + \left(\frac{\gamma_{i+1,j} - \gamma_{i,j}}{h_i} - \frac{h_i \, (\mu_{i+1,j} - \mu_{i,j})}{6}\right) \, (r - \rd{i}) \nonumber \\
    &+ \gamma_{i,j} - \frac{h_i^2 \mu_{i,j}}{6} \Bigg] \, \Theta(r - \rd{i}) \, \Theta(\rd{i+1} - r) \, ,
\end{align}
where $\mu_{i,j}$ are referred to as moments of the cubic splines, $\gamma_{i,j}$ are again given by \cref{eq:gammaij}, and $h_i \equiv \rd{i+1} - \rd{i}$.

\subsubsection*{Computation of the matrix equation}

We again evaluate the integral \cref{eq:expected_counts} and write the resulting equations in matrix form,
\begin{align}
    \begin{pmatrix} \vc{\hat{n}}_j \\ 0 \end{pmatrix}
    = \mathcal{M} \,
    \begin{pmatrix} \vc{\upgamma}_j \\ \vc{\upmu}_j \end{pmatrix} \, , \label{eq:matrix_spline}
\end{align}
where $\vc{\upgamma}_j \equiv (\gamma_{1,j}, \, \dots, \, \gamma_{\nrad,j}, \, 0)$, $\vc{\upmu}_j \equiv (\mu_{1,j}, \dots, \, \mu_{\nrad+1,j})$, $\vc{\hat{n}}_j \equiv (\hat{n}_{1,j}, \, \dots, \, \hat{n}_{\nrad,j}, \, 0 )$, and where the matrix $\mathcal{M}_{ik}$ takes the following schematic form,
\begin{equation}
\mathcal{M}
=
\begin{tikzpicture}[baseline=-0.5ex, every matrix/.style={matrix of math nodes, nodes={minimum size=6ex, inner sep=0pt, outer sep=0pt}, nodes in empty cells, left delimiter=(, right delimiter=)}]
    \matrix (m)
    {
        M_{\gamma\gamma} & M_{\gamma\mu} \\
        M_{\mu\gamma} & M_{\mu\mu} \\
    };
    \draw[red] (m-1-1.south west) -- (m-1-2.south east);
    \draw[dashed,red] (m-1-1.north east) -- (m-2-1.south east);
\end{tikzpicture}
=
\begin{tikzpicture}[baseline=-0.5ex, every matrix/.style={matrix of math nodes, nodes={minimum size=3.5ex, inner sep=0pt, outer sep=0pt}, nodes in empty cells, left delimiter=(, right delimiter=)}]
    \matrix (m)
    {
        * & * & * & \cdots & * & * & * & * & * & * & \cdots & * & * & * \\
          & * & * & \cdots & * & * & * &  & * & * & \cdots & * & * & * \\
          &  & * & \cdots & * & * & * &  &  & * & \cdots & * & * & * \\
          & & & \ddots & & & & & & & \ddots & & & \\
          &  &  &  & * & * & * &  &  &  &  & * & * & * \\
          &  &  & &  & * & * &  &  &  &  &  & * & * \\
          &  &  & &  &  & * &  &  &  & &  &  &  \\
        * & * &  & &  &  &   & * & * &  &  &  &  &  \\
        * & * & * & &  &  &   & * & * & * &  &  &  &  \\
          & * & * & \cdots &  &  &   &  & * & * & \cdots &  &  &  \\
          & & & \ddots & & & & & & & \ddots & & & \\
          &  &  & \cdots & * & * &  &  &  & * & \cdots & * & * &  \\
          &  &  &  & * & * & * &  &  &  &  & * & * & * \\
          &  &  &  &  & * & * &  &  &  &  &  & * & * \\
    };
    \draw[red] (m-7-1.south west) -- (m-7-14.south east);
    \draw[dashed,red] (m-1-7.north east) -- (m-14-7.south east);
\end{tikzpicture} \, .
\end{equation}
Here, the matrices $\mathcal{M}_{\gamma\gamma}$ and $\mathcal{M}_{\gamma\mu}$ above the solid red line come from the integral equations, while the matrices $\mathcal{M}_{\mu\gamma}$ and $\mathcal{M}_{\mu\mu}$ below the dashed red line relate to the continuity condition of the cubic spline.
We provide the concrete expressions for the $\mathcal{M}_{ik}$, and a code to carry out the related computations, on Github in the \code{SolarAxionFlux} repository~\cite{solaraxionflux}.

The only conditions needed in addition to the ones that arise from \cref{eq:expected_counts} are the boundary conditions for the cubic splines.
For example, for Hermite splines, the first derivatives are specified at the boundaries.
By choosing all of them to vanish, one obtains ``clamped'' polynomials, whose boundary conditions read,
\begin{align}
    \mathcal{M}_{\nrad,1} &= 1/h_1 \, , & \mathcal{M}_{\nrad,2} & = -1/h_1 \, , \nonumber \\
    \mathcal{M}_{\nrad,\nrad} &= h_1/3 \, , & \mathcal{M}_{\nrad,\nrad} &= h_1/6 \, , \nonumber \\
    \mathcal{M}_{2\nrad-1,\nrad-2} &= -1/h_{\nrad-1} \, ,  & \mathcal{M}_{2\nrad-1,\nrad-1} &= 1/h_{\nrad-1} \, ,  \nonumber \\
    \mathcal{M}_{2\nrad-1,2\nrad-2} &= h_{\nrad-1}/6 \, ,  & \mathcal{M}_{2\nrad-1,2\nrad-1} &= h_{\nrad-1}/3 \, .
\end{align}
Other possible options include ``natural'' boundary conditions, where the second derivatives at the boundaries are assumed to vanish.
Another widely used choice includes ``not-a-knot'' conditions, where the third derivatives of the first and last two polynomials are matched.
However, in this case the problem cannot be formulated in a straightforward matrix equation anymore and would require additional computational steps.
This also applies for other generalisation, which we discuss in \cref{app:poly_interp}

\subsubsection*{Numerical solution of the matrix equation and error propagation}

We can solve \cref{eq:matrix_spline} by numerically inverting $\mathcal{M}$ to obtain the $\gamma_{i,j}$ values needed for the fitting.
Even though the errors of the estimated observed photon counts $\hat{n}_{i,j}$ are uncorrelated, i.e.\ their covariance matrix is diagonal, the errors on the $\gamma_{i,j}$ are correlated.
If we denote the upper left block of $\mathcal{M}^{-1}$ by $M_{\gamma\gamma}^{-1}$, we find that the covariance matrix of $\vc{\upgamma}_j$ is given by
\begin{equation}
   \Sigma_j = M^{-1}_{\gamma\gamma,j} \, \mathrm{diag}( \hat{n}_{1,j}, \, \dots, \, \hat{n}_{\nrad,j}, \, 0) \, (M^{-1}_{\gamma\gamma,j})^\mathrm{T} \, . \label{eq:cubic_errors}
\end{equation}
where we again approximated the Poisson errors via $(\Delta \hat{n}_{i,j})^2 = \hat{n}_{i,j}$.
The resulting uncertainties are shown for the example in the central panel of \cref{fig:reconstruction} and can be used in \cref{eq:reconstr_fitting_metric} for the fitting procedure.

\subsection{Shape-preserving cubic spline interpolation}\label{app:poly_interp}

While instructional, there are number of issues with the direct reconstruction via matrix inversion.
Most notably the matrix needs to be non-singular.
In practical terms this means that the radial bins must be chosen such that they contain at least one (fractional) photon.
Since most of the axion flux is produced at $r \lesssim 0.25\,\Rsol$, cf.\ ref.~\cite{2101.08789}, radial bins will be concentrated in the inner region of the Sun.
\updated{As mentioned in \cref{sec:extraction}, this may lead to ``ringing'' and unphysical fluctuations below zero of the interpolating splines at large $r$.}

There exist ``shape-preserving'' spline interpolations that can guarantee properties such as positivity, convexity, or monotonicity of the resulting spline~\cite[e.g.][]{10.1137/0717021,10.1137/0905021,10.1007/BF01934097}~(see ref.~\cite{10.1145/3570157} for the included review on monotone splines).
\updated{In our study, we use the Python routine \texttt{PchipInterpolator} from the \code{scipy} library~\cite{2020SciPy-NMeth} for computing the spline coefficients for PCHIPs~\cite{10.1137/0905021} given values for \gagg, $\kappa_i$, and $T_i$.

Note that PCHIPs preserve monotonicity of the function in each interval, which does not imply that the {\GPB}s need to be monotone across all regions inside the Sun.
Due to the non-negative photon count data, monotonocity \emph{implies} positivity, which is fundamentally what we want to guarantee.
Our choice is therefore not minimal, but we argue that our method should still be viewed as model-independent.
An algorithm that only guarantees positivity would have been the most minimal choice, but software codes implementing such methods unfortunately do not seem to be widely available in \cpp or Python.

While shape-preserving splines eliminate the problem of unphysical interpolation values, one cannot write the underlying equations as a simple matrix equation anymore.
We thus have} to extend our ansatz for the {\GPB}s to arbitrary cubic polynomials.
The corresponding coefficients must then be computed by an algorithm for a given set of $\gamma_{i,j}$.
The values for $\gamma_{i,j}$ must, in turn, be computed from $\GPB_j(r_i\,|\,\gagg,\,\kappa_i,\,T_i)$.
Consequently, reconstruction of the {\GPB}s is only possible indirectly after fitting the underlying solar model parameters to the $n_{i,j}$.

More concretely, each step in the fitting procedure needs to propose values for the $\gamma_{i,j}$ coefficients, for which the adopted spline interpolation algorithm then computes the coefficients $c_{k;i,j}$ for the polynomials
\begin{equation}
    \GPB_j(r) = \sum_i \Bigg[ \gamma_{i,j} + \sum_{k=1}^{3} c_{k;i,j} (r - \rd{i})^k \Bigg] \, \Theta(r - \rd{i}) \, \Theta(\rd{i+1} - r) \, . \label{eq:general_poly}
\end{equation}
Similar to the computation in \cref{app:spline_interp}, we can compute \cref{eq:expected_counts} starting from \cref{eq:general_poly} to obtain matrix coefficients for $\mathcal{M}$ similar to \cref{eq:matrix_spline}.
However, $\mathcal{M}$ is now a $4\nrad\times\nrad$ matrix, which means that it cannot be inverted to explicitly reconstruct the {\GPB}s.
Instead, the {\GPB}s can be determined from the best-fitting points obtained by optimising \cref{eq:fitting_metric}.
For example, in the right panel of \cref{fig:reconstruction}, we show the  $\GPB_j$ and their uncertainties, assuming the same setup as in the main text (cf.\ \cref{tab:params}).

\setlength{\bibsep}{0.25ex plus 0.25ex}
\footnotesize
\renewcommand{\baselinestretch}{0.5}\footnotesize
\bibliographystyle{apsrev_mod}
\bibliography{bibliography}
\end{document}

%% file: include/journals.tex
\def\araa{ARA\&A}%
\def\apj{ApJ}%
\def\aap{A\&A}%
\def\jcap{JCAP}%
\def\mnras{MNRAS}%
\def\prd{Phys.\ Rev.\ D}%
\def\prl{Phys.\ Rev.\ Lett.}%
\def\pasp{PASP}%
\def\solphys{Sol.\ Phys.}%
\def\ssr{Space~Sci.\ Rev.}%
\def\nat{Nature}%
\def\physrep{Phys.\ Rep.}%

%% file: include/definitions.tex
\newcommand{\cpp}{C\nolinebreak\hspace{-.05em}\raisebox{.3ex}{\relsize{-2}+}\nolinebreak\hspace{-.05em}\raisebox{.3ex}{\relsize{-2}+}\xspace}

\newcommand{\code}[1]{\texttt{#1}\xspace}

\newcommand{\mtx}[2]{#1_\text{#2}}
\newcommand{\nmm}[2]{\newcommand{#1}{\ensuremath{#2}\xspace}}

\newcommand{\alphaEM}{\alpha_{\scriptscriptstyle\text{EM}}}

\newcommand{\ax}{a}
\nmm{\fax}{f_\ax}
\nmm{\ma}{m_\ax}
\nmm{\cagg}{C_{\ax\gamma\gamma}}
\nmm{\gagg}{g_{\ax\gamma\gamma}}
\nmm{\caee}{C_{{\ax}ee}}
\nmm{\gaee}{g_{{\ax}ee}}
\nmm{\Pag}{P_{\ax\gamma}}

\nmm{\PhiP}{\Phi^\text{P}}
\nmm{\GP}{\Gamma^\text{P}}
\nmm{\GPB}{\bar{\Gamma}^\text{P}}

\newcommand{\sol}{\ensuremath{\odot}\xspace}
\nmm{\Rsol}{\mathrm{R}_\sol}
\nmm{\dE}{d_\text{E}}
\nmm{\ks}{\kappa_\text{s}}
\nmm{\ompl}{\omega_\text{pl}}

\newcommand{\case}[1]{Case~#1\xspace}
\nmm{\nerg}{n_\omega}
\nmm{\nr}{n_r}
\nmm{\nrad}{n_\rho}
\nmm{\npx}{\mtx{n}{px}}
\newcommand{\om}[1]{\ensuremath{\omega_{#1}}\xspace}
\newcommand{\rh}[1]{\ensuremath{\rho_{#1}}\xspace}
\newcommand{\rd}[1]{\ensuremath{r_{#1}}\xspace}

\newcommand{\ee}{\mathrm{e}}

\newcommand{\dd}{\mathrm{d}}
\newcommand{\vc}[1]{\boldsymbol{\mathbf{#1}}}